\documentclass[a4paper]{jpconf}
\usepackage{graphicx}
\begin{document}
\title{Neutrino astrophysics : recent advances and open issues}

\author{Cristina Volpe}

\address{Astro-Particule et Cosmologie (APC), CNRS UMR 7164, Universit\'e Denis Diderot,\\ 10, rue Alice Domon et L\'eonie Duquet, 75205 Paris Cedex 13, France.}

\ead{volpe@apc.univ-paris7.fr}

\begin{abstract}
We highlight  recent advances in neutrino astrophysics,  the open issues and the interplay with neutrino properties. We emphasize the important progress  in our understanding of neutrino flavor conversion in media.  We discuss the case of solar neutrinos, of core-collapse supernova neutrinos and of SN1987A, and of the recently discovered ultra-high energy neutrinos whose origin is to be determined.
\end{abstract}

\noindent
Neutrinos are weakly interacting particles that travel over long distances and tell us properties of the environments that produce them. These elusive particles have kept mysterious for a long time. After 1998 many unknown properties have been determined thanks to the discovery of neutrino oscillations, first proposed in \cite{Pontecorvo:1957cp}. This observation was made by the Super-Kamiokande experiment using atmospheric neutrinos \cite{Fukuda:1998mi}. The neutrino oscillation discovery is fundamental for  particle physics, for astrophysics and for cosmology.

Neutrino oscillations is an interference phenomenon among the $\nu$ mass eigenstates, that occurs if neutrinos are massive and if the mass (propagation basis) and the flavor (interaction basis) do not coincide. The matrix that relates the two basis is called the Maki-Nakagawa-Sakata-Pontecorvo matrix \cite{Maki:1962mu}
which depends on three mixing angles, one Dirac  and two Majorana CP violating phases (if only three active neutrinos are considered).
Solar, reactor and accelerator experiments have determined  $\Delta m_{23}^2 = m_{3}^2 -  m_{2}^2 = 7.6 \times 10^{-3} $eV$^2$,  and $\Delta m_{12}^2 = m_{2}^2 -  m_{1}^2 = 2.4 \times 10^{-5} $eV$^2$, referred to as  the atmospheric and the solar mass-squared differences respectively.
Moreover the sign of $\Delta m^2_{12}$ has been measured by the occurrence of the Mikheev-Smirnov-Wolfenstein (MSW) effect in the Sun \cite{Wolfenstein:1977ue,Mikheev:1986gs}.
The sign of $\Delta m_{23}^2$ is still unknown which makes that there two possible ways of ordering the mass eigenstates. If $\Delta m_{31}^2 > 0$  
the lightest mass eigenstate is $m_1$ (normal ordering or "hierarchy"), whereas if  $\Delta m_{31}^2 < 0$  it is $m_3$ (inverted ordering).
Most of neutrino oscillation data can be interpreted within the framework of three active neutrinos. However a few measurements present anomalies that require further clarification. While no single hypothesis of non-standard physics can explain all anomalies, the existence of sterile neutrinos, that do not couple to the gauge bosons, can partially explain observations  (see e.g. \cite{Gariazzo:2013gua}). In neutrino oscillation experiments such neutrinos manifest themselves through the mixing with the other active species.   

Among the open questions are the mechanism for the neutrino mass, the absolute mass value and ordering, the neutrino nature (Dirac versus Majorana), the existence of CP violation in the lepton sector and of sterile neutrinos. In the coming decade(s) experiments are planned to determine some of these fundamental issues. Cosmological and astrophysical neutrinos span a wide energy range from the meV to the PeV. They are interesting both as probes of the sources that produce them and for the information they give us on neutrino themselves (see e.g. \cite{Volpe:2013kxa}). Historically astrophysical neutrinos have brought milestones in 
our knowledge of particle physics and astrophysics. They can offer avenues for future searches on key open issues.

\section{Solar neutrinos}
Electron neutrinos are constantly produced in our Sun through the proton-proton (pp) nuclear reaction chain that produces 99 $\%$ of the Sun's energy by burning hydrogen into helium-4  \cite{Bethe:1939bt}. The corresponding solar neutrino flux receives contributions from both fusion reactions and beta-decays of $^{7}$Be and $^{8}$B. First measured by R. Davis pioneering experiment \cite{Davis:1968cp}, solar neutrinos were found to be nearly a factor of three below predictions \cite{Bahcall:1968hc}. Over the decades solar neutrino experiments  have precisely measured electron neutrinos from the different pp branches, usually referred to as the pp, pep, $^{7}$Be and $^{8}$B and hep neutrinos  (Figure 1). 
The measurement of a reduced solar neutrino flux compared to standard solar model predictions -- the so-called "solar neutrino deficit problem" -- 
has been confirmed by these experiments mainly sensitive to electron neutrinos, but with some sensitivity to the other flavors. 
The advocated solutions included unknown neutrino properties (e.g. flavor oscillations, a neutrino magnetic moment coupling to the solar magnetic fields, neutrino decay, the MSW effect) and questioned the standard solar model. 
\begin{figure}
\begin{center}
\includegraphics[width=0.9\textwidth]{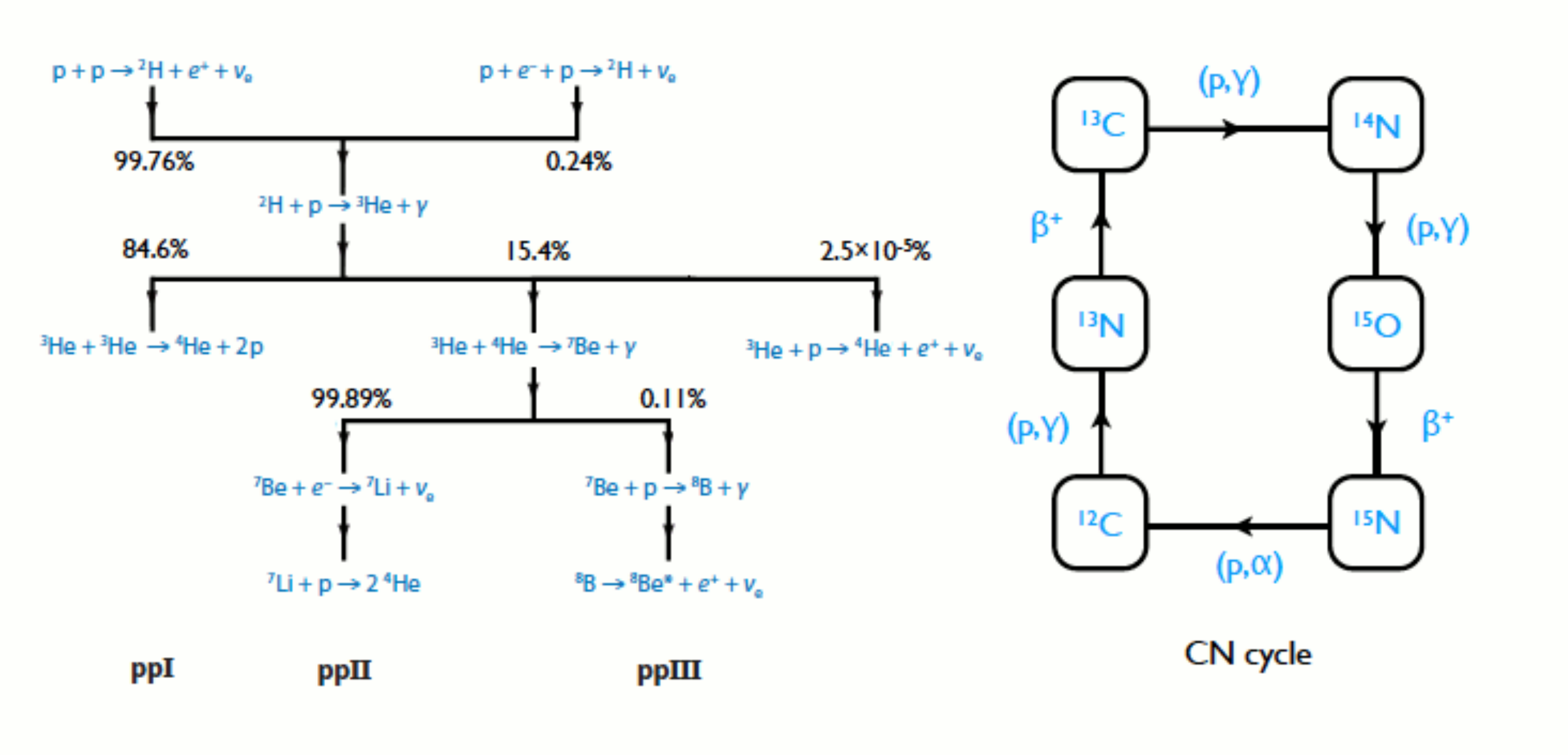}
\caption{The left figure shows  the proton-proton (pp) nuclear reaction chain with its three branches. The pp chain is responsible for energy production in our Sun and low mass stars. The theoretical branching percentages define the relative rates of the competing reactions. The right figure shows the CN cycle which is thought to play an important role for energy production in massive stars. The $^{15}$O and $^{13}$N neutrinos have not been observed yet \cite{Robertson:2012ib}.}
\label{fig1}
\end{center}
\end{figure}

The MSW effect is a resonant flavour conversion phenomenon due to the neutrino interaction with matter. 
When matter is sufficiently dilute the neutrino interaction with the medium is accounted for in the mean-field approximation.
The inclusion of the matter potential introduces a MSW resonance depending on $\nu$ properties on one hand (energy, mass-squared difference values and signs, mixing angles) and the density profile of the environment on the other.  If the evolution at resonance is adiabatic, electron neutrinos can efficiently convert into muon and tau neutrinos, thus producing a flux deficit.

The solar puzzle was solved by the discovery of the neutrino oscillation phenomenon and the results obtained by the SNO and KamLAND experiments (see \cite{Robertson:2012ib} for a review). In fact, using elastic scattering, charged- and neutral-current neutrino interactions on heavy water, the SNO experiment has showed that the measurement of the total $^{8}$B solar neutrino flux is consistent with the predictions of the standard solar model : solar electron neutrinos convert into the other active flavors. In particular, the muon and tau neutrino components of the solar flux have been measured at 5 $\sigma$ \cite{Ahmad:2002jz}. Moreover the reactor experiment KamLAND has definitely identified the Large Mixing Angle (LMA) solution, by observing reactor electron anti-neutrino disappearance at an average distance of 200 km \cite{Eguchi:2002dm}. These observations show that low energy solar neutrinos are suppressed by averaged vacuum oscillations while neutrinos having more than 2 MeV energy 
are suppressed because of the MSW effect (Figure 2). Theoretically one expects $P (\nu_e \rightarrow \nu_e)  \approx 1 - {1 \over 2} \sin^2 2 \theta_{12} \approx 0.57 $
(with $\theta_{12} = 34^{\circ}$) for ($< 2$ MeV) solar neutrinos. 
For the high energy portion of the $^{8}$B spectrum, the matter-dominated survival probability is $P (\nu_e \rightarrow \nu_e) ^{high~density}  \rightarrow \sin^2\theta_{12} \approx 0.31$ (see e.g.\cite{Robertson:2012ib}).
\begin{figure}
\begin{center}
\includegraphics[width=0.7\textwidth]{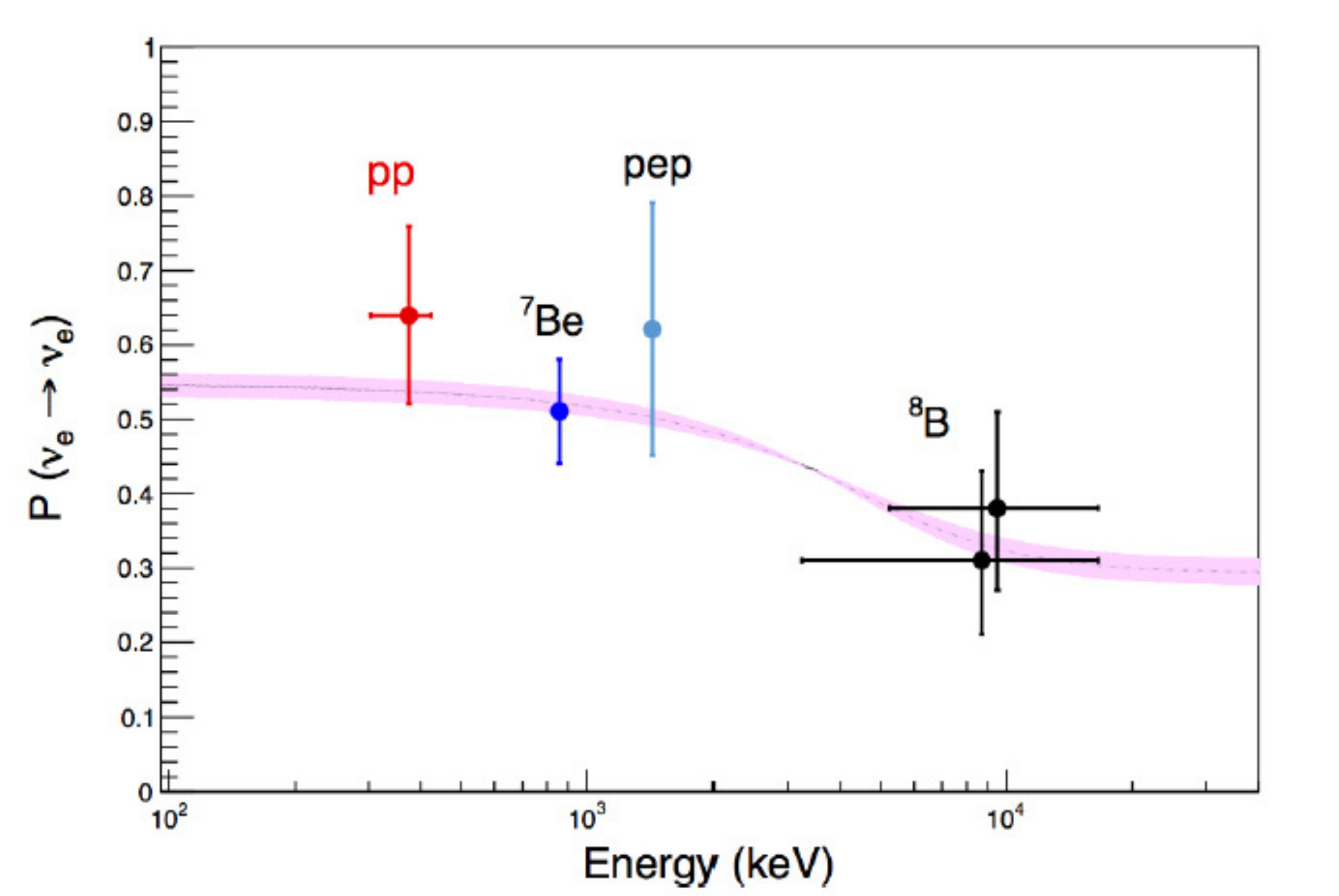}
\caption{Electron neutrino data on the solar pp, pep, $^{7}$Be, $^{8}$B neutrinos from the pp fusion chain, as a function of the neutrino energy $E_{\nu}$.  The results from the Borexino experiment are compared to the averaged vacuum oscillation prediction ($E_{\nu} < 2$ MeV) and the MSW-LMA  prediction ($E_{\nu} > 2$ MeV), taking into account present uncertainties on mixing angles \cite{Bellini:2014uqa}. }
\label{fig1}
\end{center}
\end{figure}

The Borexino experiment has precisely measured the low energy part of the solar neutrino flux, namely the pep \cite{Collaboration:2011nga}, $^{7}$Be \cite{Arpesella:2008mt} and pp neutrinos  \cite{Bellini:2014uqa}. In fact, by achieving challenging reduced backgrounds, the collaboration has just reported the first direct measurement of pp neutrinos, the keystone of the fusion process in the Sun. The measured flux is consistent with the standard solar model predictions  \cite{Bellini:2014uqa}.

The study of solar neutrinos has established that the Sun produces $3.84 \times 10^{33}$ ergs/s via the pp chain. With the measurement of pp neutrinos, it is clear that energy production in low mass stars such as our Sun is well understood.  Moreover the occurrence of the MSW effect for high energy solar neutrinos shows that 
different $\nu$ flavor conversion effects can take place in matter compared to vacuum. This phenomenon is also relevant in other contexts, including the early universe (at the epoch of the primordial abundances formation), in massive stars, in accretion disks around black holes and
in the Earth. A future goal for solar experiments is the measurement of solar neutrinos from the Carbon-Nitrogen-Oxygen (CNO) cycle which is thought to be the main mechanism for energy production in massive stars  \cite{Bethe:1939bt} (Figure 2). Borexino experiment has provided the strongest constraint on the CNO cycle that is consistent with standard solar model predictions \cite{Collaboration:2011nga}. Moreover precisely determining  the transition between the vacuum averaged and the MSW-LMA solution is of interest, since deviations from the simplest vacuum-LMA transition could point to new physics, such as non-standard interactions \cite{Friedland:2004pp}.

\section{Supernova neutrinos}
\subsection{Core-collapse supernovae and SN1987A}
Supernovae (SNe) are massive stars that undergo gravitational collapse. 
The implosion of stellar cores is at the origin of core-collapse supernovae of type II and Ib/c and of hyper-novae. It was realised in \cite{Colgate:1966ax} that a gravitational binding energy of the order of $E \approx G M_{NS}^2/R_{NS} > 10^{53}$ erg\footnote{$M_{NS}$ and $R_{NS}$ are the mass and radius of the newborn neutron star respectively} associated with the collapse of the star core to a neutron star (NS) would be released as neutrino emission and that they eventually play a role in the ejection of the stellar mantle. 
SNe are of type II if they exhibit H lines in their spectra, with type IIb having a thin H envelope. SNe II-P or II-L present a plateau or a linear decay of the light curves after the peak.    
SNe are of type I when no H lines are seen because the star has lost the H envelope. Type Ib shows He and Si lines, while Ic shows none of these indicating that the star has lost both the H envelope and He shell before collapse.  
Most of supernova light curves are due to energy deposited by the shock or to radioactivity from e.g. $^{56}$Ni decay. In some cases the radioactive component can be weak or absent. The supernova can still appear as bright if the H envelope is present, otherwise it can be invisible (Type Ib/c)  \cite{Heger:2002by}.

\begin{figure}\label{fig:2}
\begin{center}
\includegraphics[width=0.8\textwidth]{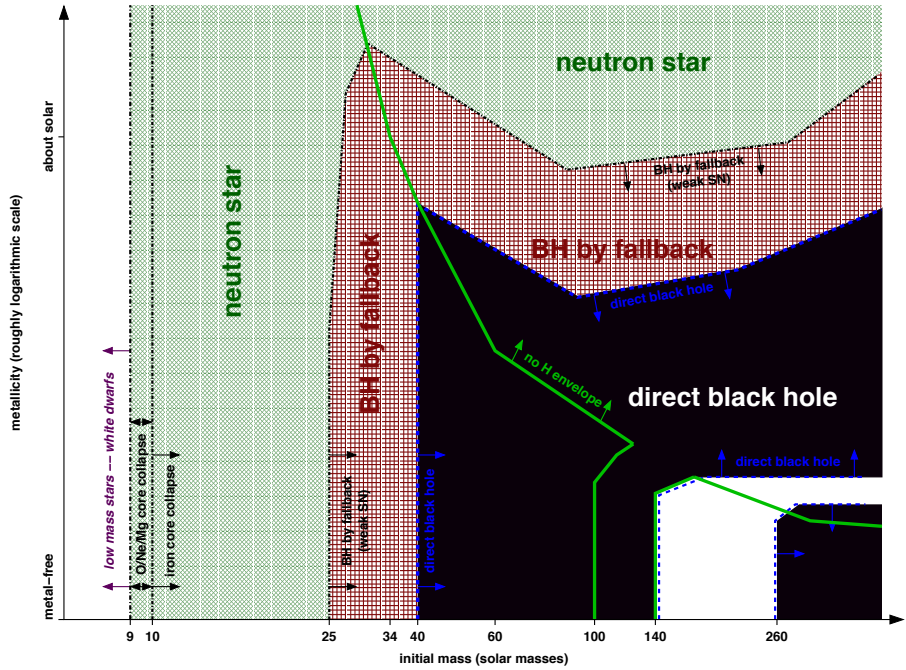}
\caption{Remnants of massive stars as a function of initial metallicity ($y$-axis) and initial mass  ($x$-axis) \cite{Heger:2002by}. }
\label{fig1}
\end{center}
\end{figure}
The fate of a massive star is mainly determined by the initial mass and composition and the history of its mass loss. The explosion produces either neutron stars or black holes, directly or by fallback (Figure 3) \cite{Heger:2002by}.
SNe initial masses range from 9 to 300 solar masses ($M_{sun}$). Stars having 9-10 $M_{sun}$ develop an O-Ne-Mg core while if $M > 10$ $M_{sun}$  an iron core is reached before collapse.  Hyper-novae are asymmetric stellar explosions with high ejecta velocities, they are very bright, producing a large amount of nickel. Hyper-novae are often associated with long-duration gamma-ray-bursts. Collapsars are all massive stars whose core collapses to a black hole and that have sufficient angular momentum to form a disk (see e.g. \cite{Heger:2002by,Janka:2012wk}).

On 23 February 1987 Sk -$69^{\circ} 202$ exploded producing SN1987A, the first naked-eye supernova since Kepler's one in 1604. It was located in the Large Magellanic Cloud, a satellite galaxy of the Milky Way. The determined distance is 50 kpc from the Earth based on the expanding photosphere method from different groups whose results agree within 10 $\%$ (see Table I of \cite{Schmidt:1992yr}). This method allows to establish extragalactic distances and cover a wide range, from 50 kpc to 200 Mpc. 
From the observed light-curve and simulations it appears that the core mass of SN1987A progenitor was around 6 $M_{sun}$ and total mass $\approx$ 18 $M_{sun}$ and the progenitor radius  about $10^{12}$ cm \cite{Pod:92}. 
SN1987A is unique because it was observed in all wavelengths from gamma rays to radio, and for the first time, neutrinos were observed from the collapse of the stellar core. These neutrinos were first discovered  by Kamiokande II \cite{Hirata:1987hu}, then by IMB \cite{Bionta:1987qt} and Baksan \cite{Alekseev:1988gp}. The number of detected electron anti-neutrinos events were 16 in Kamiokande II, 8 in IMB and 5 in Baksan. Time, energy, SN-angle  and background rate for all events are given in the nice review \cite{Vissani:2014doa}. Several hours before, 5 events were seen in LSD detector that could be due to a speculative emission phase preceding the ones seen in the other detectors  \cite{Aglietta:1987it}. Such events are often discarded in the analysis of SN1987A data since their are object of debate. 
The earliest observations of optical brightening were recorded 3 hours  after neutrinos' arrival.  An enthusiastic description of SN1987A discovery is reported in \cite{Suzuki:2008zzf}. 

\begin{figure}\label{fig:2}
\begin{center}
\includegraphics[width=0.6\textwidth]{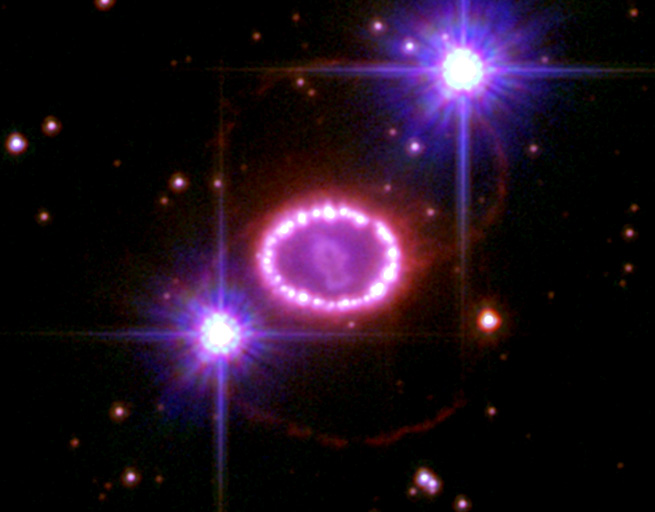}
\caption{Picture of SN1987A 20 years after its explosion with its three rings. The inner blowing ring was formed 20000 years before the explosion \cite{HTS}.}
\label{fig:sn1987}
\end{center}
\end{figure}

Three puzzling features concerning SN1987A have set constraints on stellar evolutionary models and supernova simulations. The progenitor was a blue supergiant rather than a red supergiant, while type II supernovae were thought to be produced by red supergiants. Large-mixing processes had transported radioactive nuclei from the deep core far into the H envelope of the progenitor and in the pre-supernova ejecta, producing anomalous chemical abundances. The presence of  three ring-like geometry of the circumstellar nebula around the supernova was implying a highly non-spherical structure of the progenitor envelope and its winds (Figure \ref{fig:sn1987})  \cite{Pod:92}. Various explanations have been suggested for the presence of these rings, the inner one is now dated  20000 years before the explosion. One possibility origin is a binary merger event of that epoch \cite{Pod:92, Podsiadlowski:2007zz}. In this case rotation might have played a significant role in the dying star. However the prolate deformation of the supernova ejecta at the center of the inner ring might have a very different origin (Figure \ref{fig:sn1987}).
In fact, the presence of large-mixing and of asymmetric ejecta might be suggesting the breaking of spherical symmetry due to hydrodynamical instabilities such as the bipolar Standing Accretion Shock Instability (SASI)  \cite{Janka:2007yu}.  SN1987A remnant has not been found yet. It is likely not a black hole since the progenitor was light enough to be stabilized by nuclear equation-of-states consistent with measured neutron star masses \cite{Janka:2007yu,Sato:1987yi}. There is currently no sign of a bright pulsar as well, like the one  born from the supernova explosion  in the Crab nebula in 1054.

Neutrino observations from SN1987A have been used to derive constraints on fundamental physics and the properties of neutrinos, axions, majorons, light supersymmetric particles
and on unparticles. These are derived by the absence of non-standard signatures, by using the intrinsic neutrino signal dispersion or the cooling time of the newborn neutron star (see e.g. \cite{Raffelt:1990yu}).  Many such limits have been superseded by direct measurements with controlled sources on Earth, while others remain. For example, from the three hours delay in the transit time of  neutrinos and photons, a tight limit is deduced on the difference between the neutrino speed $c_{\nu}$ and light speed $c$, i.e.  $\mid (c_{\nu} - c)/c \mid  < 2 \times 10^{-9}$  \cite{Longo:1987ub}. 

\begin{figure}\label{fig:2}
\begin{center}
\includegraphics[width=0.4\textwidth]{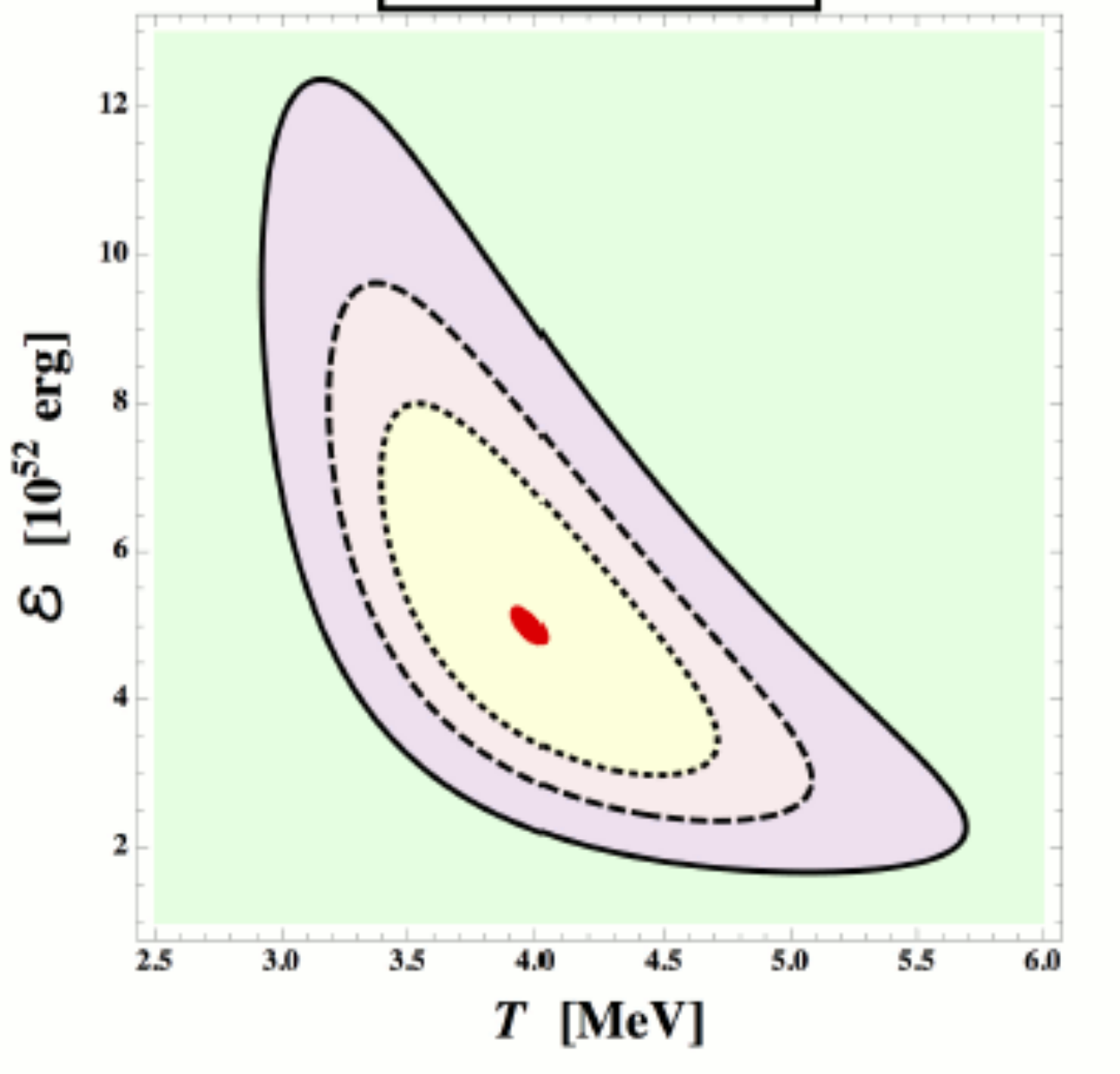}
\caption{Energy radiated in electron antineutrinos from SN1987A versus temperature of the electron anti-neutrinos. The contours show the allowed regions from a combined analysis of the 29 events recorded in the three detectors - Kamiokande II, IMB and Baksan. The confidence levels are of 1 $\%$, 68.3 $\%$, 90 $\%$ and 99 $\%$ 
\cite{Vissani:2014doa}.}
\label{fig:bestfit}
\end{center}
\end{figure}

SN1987A neutrinos have also confirmed the basic features of core-collapse supernova predictions concerning the neutrino fluence (time-integrated flux) and spectra.  From a comparative analysis of the observed neutrino events one gets as a best fit point $E = 5 \times 10^{52}$ erg and $T = 4$ MeV for the total gravitational energy radiated in electron anti-neutrinos and their temperature respectively (Figure \ref{fig:bestfit}). The emission time is also found to be of $15$ s \cite{Vissani:2014doa}.
According to expectations, 99 $\%$ of the supernova gravitational binding energy should be converted in $\nu_e, \nu_{\mu}, \nu_{\tau}$ neutrinos (and anti-neutrinos) in the several tens of MeV energy range. If one considers that energy equipartition among the neutrino flavors is rather well satisfied, one gets about $3 \times 10^{53}$ ergs, in agreement with expectations. 
Moreover, the neutrino spectra are thermal to a fairly good approximation. In this case the average electron anti-neutrino energy  is related to the temperature by $E_{\nu} = 3$ $T$, giving 12 MeV at the best fit point. For a long time this value has appeared much lower than the expected value of about $15$ MeV.
Currently  it appears rather compatible with supernova simulations based on realistic neutrino transport.
 
Supernova neutrinos are tightly connected with two major questions in astrophysics, namely what is the mechanism that makes massive stars explode and what are the sites where the heavy elements are formed. 
Besides the astrophysical conditions and the properties of exotic nuclei, neutrinos also determine 
the heavy element nucleosynthetic abundances. In fact, the interaction of electron neutrinos and anti-neutrinos with neutrons and protons in such environments determines the neutron-to-proton ratio, a key parameter of the r-process. Among the candidate sites for heavy elements nucleosynthesis are core-collapse supernovae, accretion-disks around black holes and neutron-star mergers. Numerous studies show that neutrinos impact the neutron richness of a given astrophysical environment. In order to finally assess the neutrino impact, extensive simulations need still to be performed, which e.g. self-consistently determine neutrino and matter evolution,  (see the reviews in the Focus Issue \cite{Volpe:2014yqa}). 

Various mechanisms for the SN blast have been investigated in the last decades, including thermonuclear, bounce-shock, neutrino-heating,  magnetohydrodynamic, acoustic and phase-transition mechanisms (see  \cite{Janka:2012wk}). Since the kinetic energy in SN events goes from $10^{50-51}$ erg for SNe up to several $10^{52}$ erg for hyper-novae, the explosion driving mechanism have, among others, to provide such energies. The neutrino-heating mechanism with non-radial hydrodynamical instabilities (convective overturn with SASI) appears a good candidate to drive iron-core collapse supernova explosions; while the more energetic hypernovae events could be driven by the magnetohydrodynamical mechanism  \cite{Janka:2012wk}. Note that a new neutrino-hydrodynamical instability  termed LESA  (Lepton-number Emission Self-sustained Asymmetry) has been identified  \cite{Tamborra:2014aua}.
Successful explosions  for two-dimensional simulations with realistic neutrino transport have been obtained for several progenitors; while the first  three-dimensional explosion has just been obtained which shows an enhanced expansion compared to two-dimensional simulations. This results from a combination of neutrino heating, SASI and  turbulent convection \cite{Melson:2015tia}.

 \subsection{Neutrino flavor conversion in massive stars} 
Important progress has been achieved in understanding how neutrinos change their flavor in massive stars. Besides the MSW effect \cite{Dighe:1999bi}, calculations have revealed new phenomena due to the neutrino-neutrino interaction, the presence of shock-waves and of turbulence (see \cite{Duan:2010bg,Duan:2009cd} for a review).  In particular steep changes of the star density profile due to shock-waves induce multiple MSW resonances and interference among the matter eigenstates \cite{Dasgupta:2005wn,Kneller:2007kg}. As a consequence the neutrino evolution can become completely non-adiabatic when the shock passes through the MSW region.

The neutrino evolution in flavor space can be understood using the correspondence with spin dynamics. In this framework
the Schr\"odinger-like equation that governs the evolution of the neutrino flavor amplitudes is replaced by the precession for spins submitted to effective magnetic fields (see e.g. \cite{Cohen}).
The components of the magnetic field are the real and imaginary part of the off-diagonal term of the Hamiltonian and the difference of the diagonal terms. The third component of the spins gives the neutrino flavor content while the other two depend on the neutrino mixings. 

Ref.  \cite{Pantaleone:1992eq} first pointed out that the inclusion of the neutrino-neutrino interaction introduces a non-linear refractive index.
Significant effects have been found in \cite{Balantekin:2004ug,Duan:2005cp} which has triggered intensive investigations since. It has emerged that
the neutrino self-interaction  produces  collective stable and unstable modes of the (anti-)neutrino gas. 
These findings are based on the so-called "bulb model", 
that assumes spherical and azimuthal symmetry of the neutrino emission and propagation. 
Three flavor conversion regimes are found : synchronization, bipolar oscillations and spectral split (Figure 6). 
Nearby the neutrino-sphere the neutrino self-interaction is so strong that the individual spins are stick together and collectively precess around the neutrino self-interaction term. During synchronization no flavor conversion takes place.  However the corresponding interaction strength decreases with distance $r$ as $1/r^{4}$ \cite{Duan:2010bg}. 
At some point of the evolution, the presence of non-zero mixing triggers a flavor instability after which bipolar oscillations occur that can be seen either as a pendulum or a gyroscopic pendulum in flavor space \cite{Hannestad:2006nj}. In the matter eigenstate basis one can show that the bipolar instability is associated with the rapid growth of  the matter phase (that depends upon the imaginary components of the neutrino self-interaction)   \cite{Galais:2011jh}. After bipolar oscillations, a spectral split phenomenon produces either full or no conversion depending on the neutrino energy. This is a MSW-effect in a co-moving frame \cite{Raffelt:2007xt}, or analogous to a magnetic-resonance phenomenon  \cite{Galais:2011gh}.  It is analogous because the z-axis component of the magnetic field is not constant as is the case in the magnetic resonance phenomenon but slowly varies.  
The collective neutrino modes just described produce a swapping of the neutrino fluxes  with sharp spectral changes (Figure 7).
These are flavor conversion mechanisms that emerge  in presence of the neutrino self-interaction within the "bulb model".  The neutrino-neutrino interaction produces interesting conversion effects also in accretion disks arising from neutron star-neutron star  or black hole-neutron star mergers \cite{Malkus:2014iqa}. Note that numerous other oscillation effects have been studied, such as spin-flavour oscillations of neutrinos in the generation of pulsar kicks in the gravitation field of magnetized neutron stars \cite{Lambiase:2004qk}. 
\begin{figure}[h]
\begin{minipage}{20pc}
\includegraphics[width=17pc]{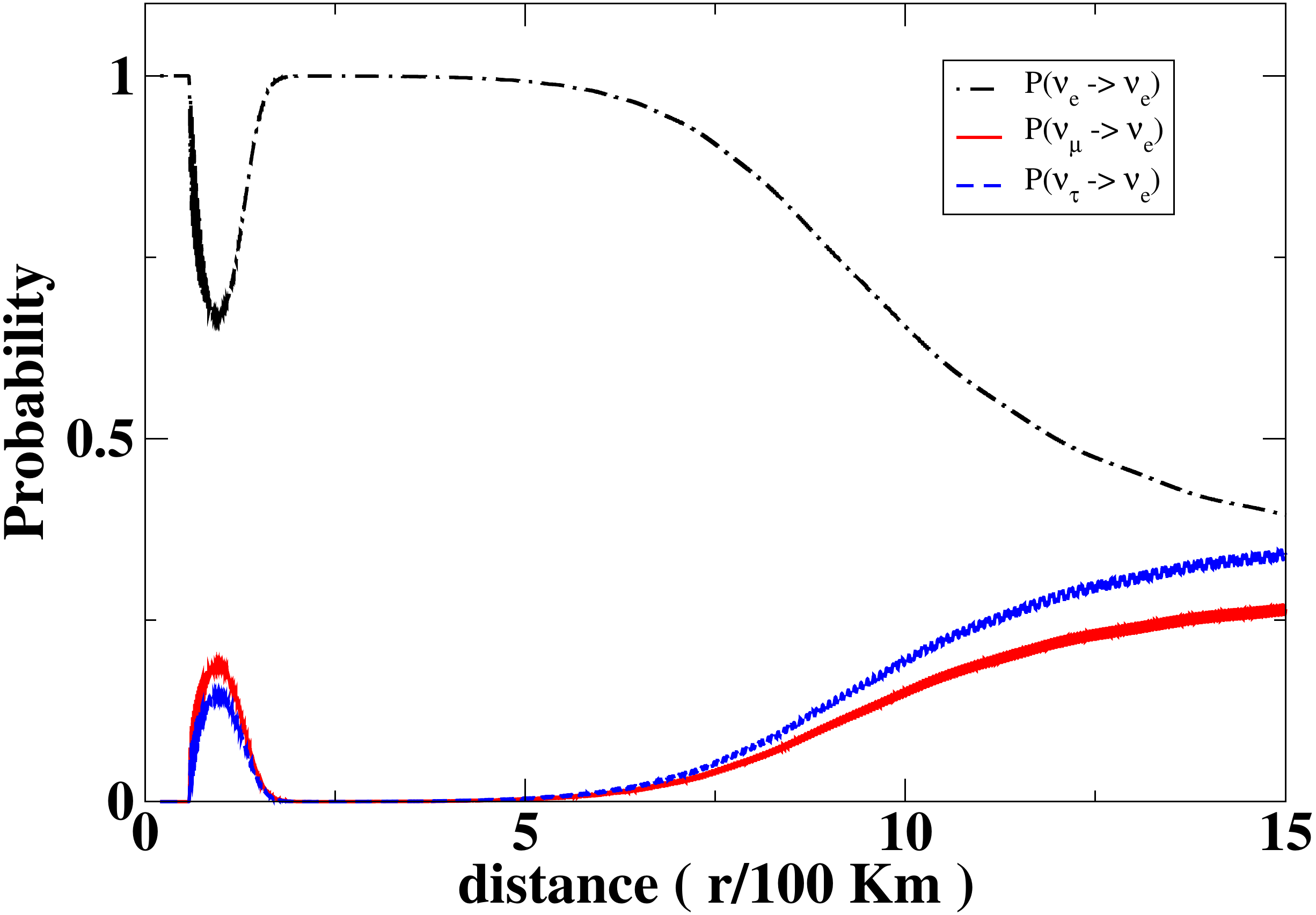}
\caption{\label{label} Electron neutrino survival probability in  a core-collapse supernova about 1s after bounce. The neutrino-sphere is taken at a neutron star radius of 10 km. The simulation includes the neutrino mixing, the neutrino-matter and neutrino-neutrino interaction in the single angle approximation. The results for a 5 MeV neutrino are for inverted hierarchy. Synchronization, bipolar oscillations and spectral split can be seen in the first 200 km  (see text) \cite{Gava:2008rp}. }
\end{minipage}\hspace{2pc}%
\begin{minipage}{20pc}
\includegraphics[width=17pc]{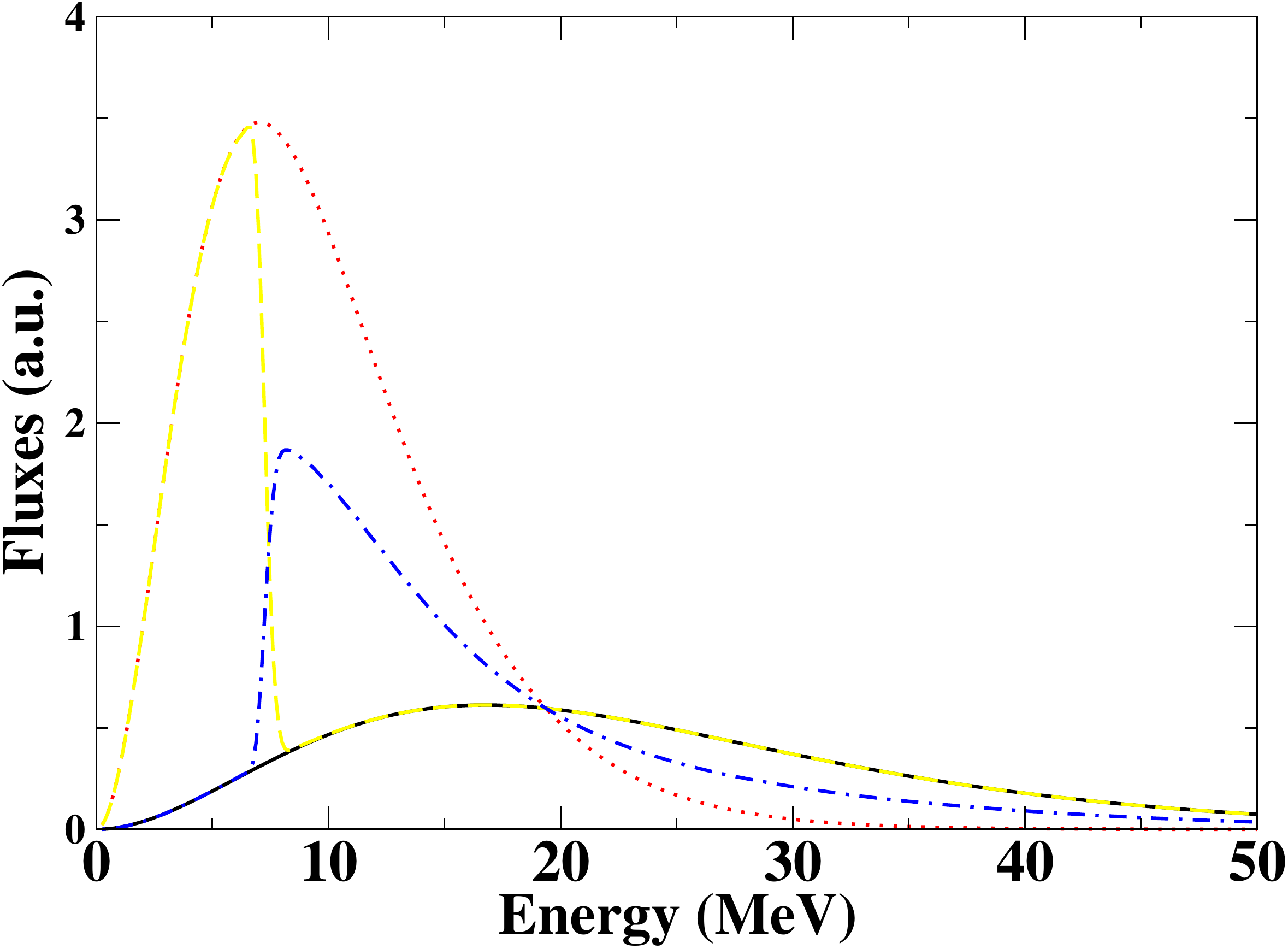}
\caption{\label{label}Neutrino fluxes in  a core-collapse supernova about 1s after bounce. The red-dotted (black) line corresponds to electron (muon and tau) neutrinos
at the neutrino-sphere, while the yellow (blue dot-dashed) line is the corresponding one after the neutrino-neutrino interaction effects.
The results are at 200 km from  the neutrino-sphere (at 10 km). The simulation is the same as for Figure 6. Due to the neutrino-neutrino interaction, a swapping between the electron and the non-electron neutrino types takes place at a characteristic energy, here at around 8 MeV  \cite{Gava:2008rp}.}
\end{minipage} 
\end{figure}

Recent investigations have clearly shown that going beyond approximations inherent to the "bulb" model introduces a higher degree of complexity of the flavor evolution. 
First of all an assumption often made within the bulb model is that neutrinos are all emitted at the neutrino-sphere with the same angle ("single-angle" approximation), thus maximizing the coherence of the modes.
However the interplay between matter and neutrino self-interaction effects need to be accurately considered. In fact in the so-called "multi-angle simulations" when the matter density exceeds  or is
comparable to the neutrino density, collective effects can be strongly suppressed since neutrinos with different emission angles have different flavor histories \cite{EstebanPretel:2008ni}.  In particular it appears as for now that 
collective effects are suppressed during the accretion phase in  simulations based on realistic density profiles from supernova one-dimensional simulations \cite{Chakraborty:2011gd}. If this result is maintained in future calculations it implies that during the neutronization burst (50 ms after bounce) and the accretion phase (several hundreds ms after bounce) neutrinos change their flavor only due to the well understood MSW effect in the outer layers. Note that shock-wave effects impact the neutrino fluxes only after about 1 s after bounce in the anti-neutrino (neutrino) sector if the neutrino mass ordering is inverted (normal). 
During the neutronization burst neutrino self-interaction effects are absent because, they imply $\nu_e \bar{\nu}_e \rightarrow \nu_x \bar{\nu}_x $. In fact,  at an early stage of the collapse $\nu_e$ only are emitted by electron capture on nuclei.

Besides the bipolar,  other flavor instabilities appear by spontaneous breaking of the symmetries of  the "bulb model", such as the azymuthal symmetry \cite{Raffelt:2013rqa}. This symmetry breaking appears even if the initial conditions and the evolution equations have such a symmetry. 
This has been shown using linearised analysis,  first proposed in this context in \cite{Banerjee:2011fj} and futher highlighted in \cite{Volpe:2013uxl} in a broader context. 
In the small amplitude approximation (also called Random-Phase-Approximation in the many-body context) unstable (or runaway) solutions are revealed by the presence of complex eigenvalues of the stability matrix. This signals a change in the curvature of the energy density of the system around the starting point, given by the initial conditions at the neutrino-sphere. Similarly collective stable (or unstable) modes can be identified in many-body systems such as atomic nuclei and metallic clusters \cite{Volpe:2013uxl}. 

The current theoretical description of neutrino propagation in massive stars is based on the mean-field approximation. As initial conditions for the neutrino fluxes at the neutrino-sphere, results from supernova simulations are taken. The latter is assumed a sharp boundary.
For example, few collisions outside the neutrino-sphere might modify the picture. Their inclusion in a toy model has revealed significant modifications of the flavor patterns \cite{Cherry:2012zw}.

Extended descriptions describing neutrino evolution in dense media have been derived using a coherent-state path integral \cite{Balantekin:2006tg}, the Born-Bogoliubov-Green-Kirkwood-Yvon hierarchy \cite{Volpe:2013uxl}, or the two-particle-irreducible effective action formalism \cite{Vlasenko:2013fja} (see also \cite{Sigl:1992fn,McKellar:1992ja}). 
These approaches provide a derivation of the evolution equations from first principles and using quantum field theory\footnote{Note that the quantum field theory treatment of neutrino mixing has been extensively investigated (see e.g. \cite{Blasone:1995zc,Giunti:2003dg}). The possible contribution to the cosmological constant from neutrino mixing is pointed out in \cite{Blasone:2004yh}.}. 
Besides collisions, extended mean-field descriptions present two kinds of corrections : spin or helicity coherence \cite{Vlasenko:2013fja}  and neutrino-antineutrino pairing correlations \cite{Volpe:2013uxl}. The former are present because of the neutrino mass, while the latter appear in an extended mean-field approximation when considering all possible two-point correlators. The most general equations for neutrino propagation including both corrections have been derived in \cite{Serreau:2014cfa}. While the origin is very different, both kinds of correlations introduce neutrino-antineutrino mixing, if spatial anisotropies of the matter and/or neutrino backgrounds are present. Such corrections are expected to be tiny, but the non-linearity of the equations could introduce significant changes of neutrino evolution in particular in the transition region. This is between the dense region within the neutrino-sphere which is Boltzmann treated, to the diluted one outside the neutrino-sphere where collective flavor conversion occurs. 
Numerical calculations are still needed to investigate the role of spin coherence or neutrino-antineutrino pairing correlations or of collisions.
A first calculation in a simplified model shows that helicity coherence might have an impact \cite{Vlasenko:2014bva}. 

Another interesting theoretical development is the establishment of connections between neutrino flavor conversion in massive stars and the dynamics, or behavior, of many-body systems in other domains.
Using algebraic methods, Ref.\cite{Pehlivan:2011hp} has shown that the neutrino-neutrino interaction Hamiltonian can be rewritten as a (reduced) 
Bardeen-Cooper-Schrieffer (BCS) Hamiltonian for superconductivity \cite{Bardeen:1957mv}. As mentioned above, Ref.\cite{Volpe:2013uxl} has included neutrino-antineutrino correlations of the pairing type which are formally analogous to the BCS correlations. The linearisation of the corresponding neutrino evolution equations  has highlighted the formal link between stable and unstable collective neutrino modes and those in atomic nuclei and metallic clusters \cite{Vaananen:2013qja}.  

Significant progress has been performed in unravelling how neutrinos change flavor in massive stars. However important questions require further investigations, including e.g. the assessment of their final impact on a neutrino signal on Earth, of the effect of a wave-packet treatment \cite{Akhmedov:2014ssa} or of corrections beyond the mean-field approximation \cite{Balantekin:2006tg,Volpe:2013uxl,Vlasenko:2013fja}, or the implementation of realistic matter density profiles in multidimensional simulations and of turbulence.

\subsection{Supernova neutrino observations}
The SuperNova Early Warning System (SNEWS) \cite{Antonioli:2004zb} and numerous other neutrino detectors around the world can serve as supernova neutrino observatories if a supernova blows up in the Milky Way, or outside our galaxy. Large scale detectors based on different technologies \cite{Autiero:2007zj,Scholberg:2012id} including liquid argon, water Cherenkov and scintillator are being considered (e.g. JUNO or Hyper-K). These have the potential to detect neutrinos from a galactic or an extragalactic explosion as well as to discover the diffuse supernova neutrino background produced from supernova explosions up to cosmological redshift of 2 (for a review see \cite{Beacom:2010kk,Lunardini:2010ab}).

The observation of the neutrino luminosity curve from a future (extra)galactic supernova would closely follow different explosion phases furnishing a crucial test of
supernova simulations, and  information on unknown neutrino properties. In particular, the occurrence of the MSW effect in the outer layers of the star
and of collective effects depends on the value of the third neutrino mixing angle and the neutrino mass ordering. The precise measurement of the last mixing angle \cite{Abe:2011fz,An:2012eh,Ahn:2012nd} reduces the number of unknowns. Still, the neutrino signal from a future supernova explosion could tell us about the mass ordering, either from  the early time signal in IceCube \cite{Serpico:2011ir},  or by measuring, in Cherenkov or scintillator detectors, the positron time and energy signal associated with the passage of the shock-wave in the MSW region \cite{Gava:2009pj}. Several other properties can impact the neutrino fluxes such as the neutrino magnetic moment \cite{deGouvea:2012hg}, non-standard interactions, sterile neutrinos. CP violation effects from the Dirac phase exist but appear to be small as established in \cite{Balantekin:2007es} and further studied in \cite{Gava:2008rp,Pehlivan:2014zua,Kneller:2009vd}. In spite of the range of predictions, the combination of future observations from different  detection channels can bring key information to this domain (see e.g. \cite{Vaananen:2011bf}). 

\begin{figure}
\begin{center}
\includegraphics[width=0.7\textwidth]{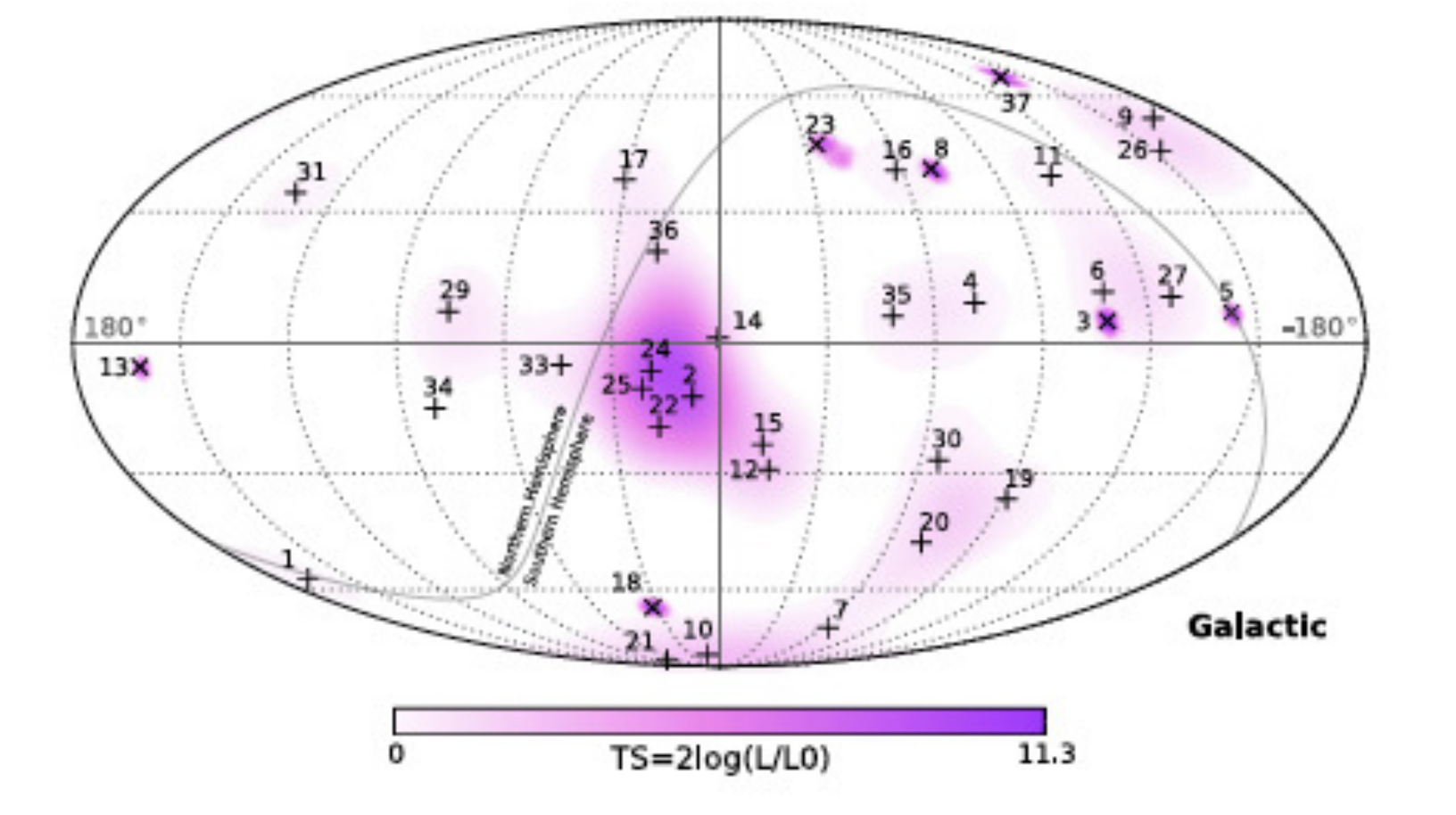}
\caption{Arrival directions of the neutrino events seen in the IceCube detectors in Galactic coordinates. The grey line shows the equatorial plane. Showerlike events  are marked with + and muon tracks events  with x. A test statistic (TS) for point source clustering at each location is shown by the colours. No significant clustering is observed \cite{Aartsen:2014gkd}.  }
\label{fig:icecube}
\end{center}
\end{figure}

\section{Ultra-high energy neutrinos}
The main mission of  high-energy neutrino telescopes is to search for galactic and extra-galactic sources of high-energy neutrinos to elucidate the source of cosmic rays and the astrophysical mechanisms that produce them. These telescopes also investigate neutrino oscillations, dark matter and supernova neutrinos (for IceCube). The 37 events collected in IceCube, with deposited energies ranging from 30 TeV to 2 PeV,  is consistent with the discovery of high energy astrophysical neutrinos at 5.7 $ \sigma$ (Figure \ref{fig:icecube})  \cite{Aartsen:2014gkd}.  The 2 PeV event is the highest-energy neutrino ever observed.

High-energy neutrino telescopes are currently also providing data on neutrino oscillations measuring atmospheric neutrinos, commonly a background for astrophysical neutrino searches.   Using low energy samples, both ANTARES \cite{AdrianMartinez:2012ph} and IceCube/DeepCore \cite{Gross:2013iq} have measured the parameters $\theta_{23}$ and $ \Delta m^2_{23}$ in good agreement with existing data. 
Neutrino telescope are also sensitive to other fundamental properties such as Lorentz and CPT violation  \cite{Abbasi:2010kx}, or sterile neutrinos. 
PINGU, IceCube extension in the 10 GeV energy range, could measure the mass hierarchy and be sensitive to the Dirac phase \cite{Akhmedov:2012ah}. Such measurements exploit the occurrence of the matter effect from neutrinos, both from the MSW and the parametric resonance occurring in the Earth \cite{Akhmedov:1998ui,Petcov:1998su}. Feasibility studies are currently ongoing both for PINGU and for ORCA  \cite{Kouchner:2014epa} to establish if the energy and angular resolution required for the mass hierarchy search can be achieved.

\end{document}